\documentclass[prb,twocolumn,amsmath,amssymb,floatfix,superscriptaddress]{revtex4-1}
\usepackage{graphicx}
\usepackage[FIGTOPCAP]{subfigure}
\usepackage[usenames]{color}
\usepackage{hyperref}
\usepackage{bbm}
\usepackage{bbold}

\begin{document}
\title{Hybridization-induced interface states in a topological-insulator--ferromagnetic-metal heterostructure}
\date{\today}
\author{Yi-Ting Hsu}
\affiliation{Condensed Matter Theory Center and Joint Quantum Institute, Department of Physics, University of Maryland, College Park, Maryland 20742}
\author{Kyungwha Park}
\affiliation{Department of Physics, Virginia Tech, Blacksburg, Virginia 24061, USA}
\author{Eun-Ah Kim}
\affiliation{Department of Physics, Cornell University, Ithaca, New York 14853, USA}

\begin{abstract}
Recent experiments demonstrating large spin-transfer torques in topological insulator (TI)-ferromagnetic metal (FM) bilayers have generated a great deal of excitement due to their potential applications in spintronics.
The source of the observed spin-transfer torque, however, remains unclear.
This is because the large charge transfer from the FM to TI layer would prevent the Dirac cone at the interface from being anywhere near the Fermi level to contribute to the observed spin-transfer torque.
Moreover, there is yet little understanding of the impact on the Dirac cone at the interface from the metallic bands overlapping in energy and momentum, where strong hybridization could take place.
Here, we build a simple microscopic model and perform first-principles-based simulations
for such a TI-FM heterostructure, considering the strong hybridization and charge transfer effects.
We find that the original Dirac cone is destroyed by the hybridization as expected.
Instead, we find a new interface state which we dub 'descendent state' to form near the Fermi level due to the strong hybridization with the FM states at the same momentum.
Such a `descendent state' carries a sizable weight of the original Dirac interface state, and thus inherits the localization at the interface and the same Rashba-type spin-momentum locking.
We propose that the `descendent state' may be an important source of the experimentally observed large spin-transfer torque in the TI-FM heterostructure.
\end{abstract}

\maketitle
\section{Introduction}
Topological insulator (TI)-based heterostructures have become appealing candidates for spintronics due to the Dirac surface states
which exhibit the spin-momentum locking with opposite Rashba spin-windings at opposite surfaces\cite{ExpTIspinwind}. In particular, large spin-transfer torques comparable to conventional heavy-metal-based structures have been reported in three-dimensional (3D) TI-based bilayers\cite{DanExp,ExpUCLAtorque,ExptorqueTICoFeB}.
Although large signals from the spin-momentum-locked Dirac surface states on the interface have been predicted in TI-ferromagnetic insulator heterostructures\cite{TIFItorque}, the materials involved in experiments are often ferromagnetic metals (FMs)\cite{DanExp,ExptorqueTICoFeB} since ferromagnetic insulators
are rare. Data analyses attributed the observed spin-transfer torques to an interface state with the same spin-winding as the Dirac surface state, when a pristine TI is in contact with a ferromagnetic layer\cite{DanExp,MarkTorque}. The identity of this state, however, remains elusive.
This is because this Dirac surface state is very likely to be buried way below the raised chemical potential or even destroyed when hybridized with a large amount of FM states under an Ohmic contact\cite{ThyTIFM}.\\

The effect of heterostructure formation on the primitive Dirac surface state localized at the interface (hence-forth referred to as the `Dirac interface state') has been investigated for various materials.
In the cases of TI-insulator bilayers, first-principles studies have found that
interface states localized a bit deeper into the interface appear near the Fermi level $E_F$ with Dirac-like dispersion, while the original Dirac surface states are shifted way below $E_F$ due to band-bending potentials, which are induced by the mismatch between chemical potentials
\cite{TIFIKyungwha,TIInsulThy,BiBi2Se3DFT,PARK13}.
As for TI-metal heterostructures, effective model studies have found Dirac interface states becoming diffusive under weak TI-metal coupling\cite{TImetalDasSarma} while first-principles studies have reported no spin-momentum-locked interface states for various metals\cite{TImetalDFT,ThyTIFM}.
In particular, severe hybridization is expected for cases where electrons in the Dirac interface states are coupled to many itinerant electrons with similar momenta and energies from a metal slab with much higher chemical potential.
Such a scenario where itinerant electrons couple with a localized state has been generically described by the Fano-Anderson model to the lowest order\cite{Mahan,Fano,TISBhyb}. Fano coupling involving enough extended states can produce a new long-lived localized state sitting outside of the
metal band which carries a substantial weight of the original localized state\cite{Mahan}.
This `descendent state' suggests the identity of the interface state with the same spin-winding as the Dirac interface state which leads to the large spin-transfer torque probed in the TI-FM heterostructure.

Our goal is to study the fate of the Dirac interface state in contact with an FM slab that has a much higher $E_F$ and many states overlapping with the Dirac interface state in energy and momentum.
In this article, we take two complementary approaches: we
construct a simple microscopic model and perform first-principles calculations, including the hybridization
between the metal states and the Dirac cone in such a TI-FM bilayer.
By examining spectroscopic properties, we identify the new interface state near $E_F$ as the
descendent state, which is localized slightly deeper into the TI layer and inherits the spin texture of the original Dirac cone. The features obtained from the microscopic model agree with those from the first-principles-based simulations of the TI-FM bilayer.
We propose that the descendent states may be an important source of the recently observed large spin-transfer torque in $\rm{Bi_2Se_3}$-$\rm{Py}$ heterostructure\cite{DanExp}.
The article is structured as follows: In Sec. II, we construct a lattice model for a TI-FM heterostructure with hybridization in the spirit of Fano-Anderson model.
In Sec. III, we study the spectroscopic properties of the model and the properties of the newly formed interface states. In Secs. IV and V, we present a first-principles study on $\rm{Bi_2Se_3}$-$\rm{Ni}$ bilayer using density functional theory (DFT) and compare the results to those from
our lattice model. In Sec. VI, we summarize our results and address open questions.

\section{Lattice model for a TI-FM bilayer in the presence of hybridization}

Generically the coupling between a localized state $f$ with energy $\epsilon^0$ and itinerant electrons $c_k$ with energy $E_k$ can be described by Fano model\cite{Mahan}
$H_F= \epsilon^0f^{\dagger}f+\sum_kE^0_kc^{\dagger}_{k}c_{k}+g_k(c^{\dagger}_{k}f+f^{\dagger}c_{k})$, where $g_k$ is the coupling strength.
\textcite{Fano} pointed out that $H_F$ can be used to model the lowest order interaction between a helical surface state and a metal bulk band, i.e. hybridization.
As shown in Ref. \onlinecite{Mahan} and \onlinecite{Fano}, the hybridized spectral function of the surface state $f$ contains a broadened peak in the metallic band and two new delta functions with a reduced weight above and below the edges of the metallic band. 
The delta functions represent a part of the spectral weight associated with the original surface state that got pushed away from the metallic band through level repulsion as a result of coupling to extended states. These states form a new long-lived surface states albeit with smaller spectral weight. 
We thus dub the new surface states the `descendent states' of the mother surface state.
As the hybridization strength $g_k$ increases, the lifetime of the original surface state shortens while both the weights of the mother state carried by the descendent states and the gap to the metal band edges increase.

To examine the formation of descendent states in a TI-FM bilayer, we construct a tight-binding model capturing the coupling between the Dirac interface state with the extended states in the spirit of Fano model. 
The model for an $(N_{TI}+N_{FM})$-layer heterostructure reads $H_0=\sum_{\mathbf{k_{\parallel}}}H_0(\mathbf{k_{\parallel}})$ with
\begin{align}
H_0(\mathbf{k_{\parallel}})=&\sum_{j=1}^{N_{TI}} H_{TI}(\mathbf{k_{\parallel}},j)+\sum_{j=N_{TI}+1}^{N_{TI}+N_{FM}}H_{FM}(\mathbf{k_{\parallel}},j),
\label{eq:Hfull}
\end{align}
where $\mathbf{k_{\parallel}}=(k_x,k_y)$ is the in-plane momentum, $j$ labels the layers stacked in the $z$ direction,
$H_{TI}$ and $H_{FM}$ are Hamiltonians for a $N_{TI}$-layer TI slab and a $N_{FM}$-layer FM slab stacked in the $z$ direction.

$H_{TI}(\mathbf{k_{\parallel}},j)$ is a four-band microscopic model describing a quintuple layer (QL) $j$ of pristine TI\cite{TISBhyb} in the presence of a band-bending potential created by the mismatch between the Fermi levels of the TI and FM:
\begin{equation}
H_{TI}(\mathbf{k_{\parallel}},j)=(\mathbb{M}+\mathbb{V})~c^{\dagger}_{\mathbf{k_{\parallel}},j}c_{\mathbf{k_{\parallel}},j}
+\mathbb{T}c^{\dagger}_{\mathbf{k_{\parallel}},j+1}c_{\mathbf{k_{\parallel}},j}
+H.c. ,
\label{eq:HTI}
\end{equation}
where $\mathbb{M}$, $\mathbb{T}$ and $\mathbb{V}$ are $4\times 4$ matrices in the basis of $|\uparrow,P_1\rangle$, $|\uparrow,P_2\rangle$, $|\downarrow,P_1\rangle$, and $|\downarrow,P_2\rangle$ with $\uparrow$/$\downarrow$ being spin and $P_{1/2}$ being $p_z$ orbitals of Bi/Se atoms.
Here $\mathbb{M}$ and $\mathbb{T}$ contain tight-binding parameterization for pure $\rm{Bi_2Se_3}$ \cite{TISBhyb,3DTInphys}, while $\mathbb{V}=V(\mathbf{k_{\parallel}},j)\mathbb{I}_{4\times 4}$ is the potential well which forms near the interface when the TI has a lower Fermi level than the FM.
For simplicity, we assume the potential to be momentum $\mathbf{k_{\parallel}}$-independent and has a spatial profile of $V(j)=V_0e^{-\eta (N_{TI}-j)}$ based on the potential shapes in various TI-based bilayers obtained in ab initio calculations\cite{TImetalDFT,PARK13}.
The depth of the well $V_0$ is approximately the Fermi level-difference between the two materials.
For our purpose of demonstrating the hybridization effect on the Dirac interface state, we choose the width $1/\eta$ to be small enough such that no additional quantum well state forms.

\begin{figure}[t]
\subfigure[]{
	\includegraphics[width=8cm]{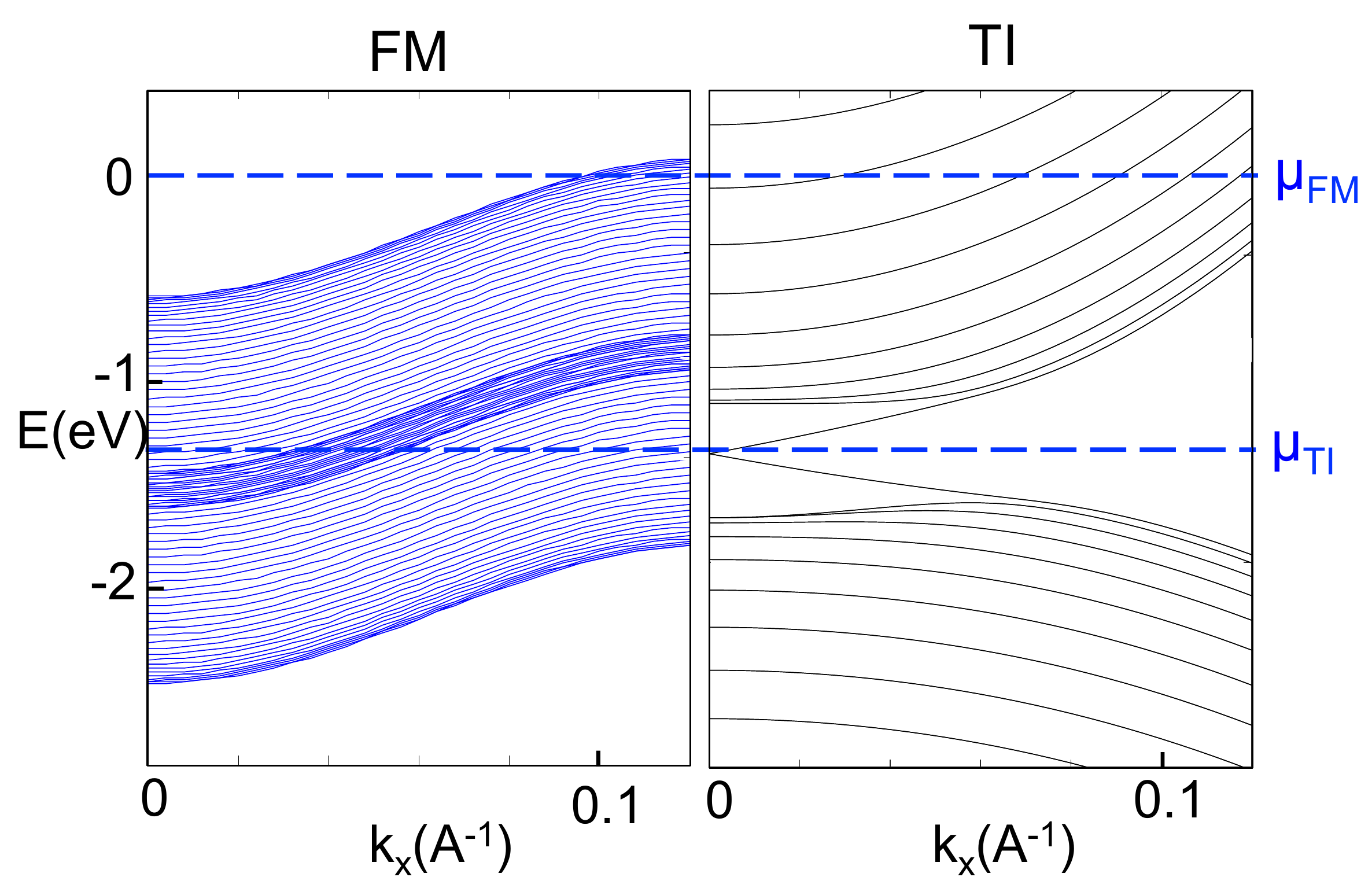}
	}
	\subfigure[]{
	\includegraphics[width=4.5cm]{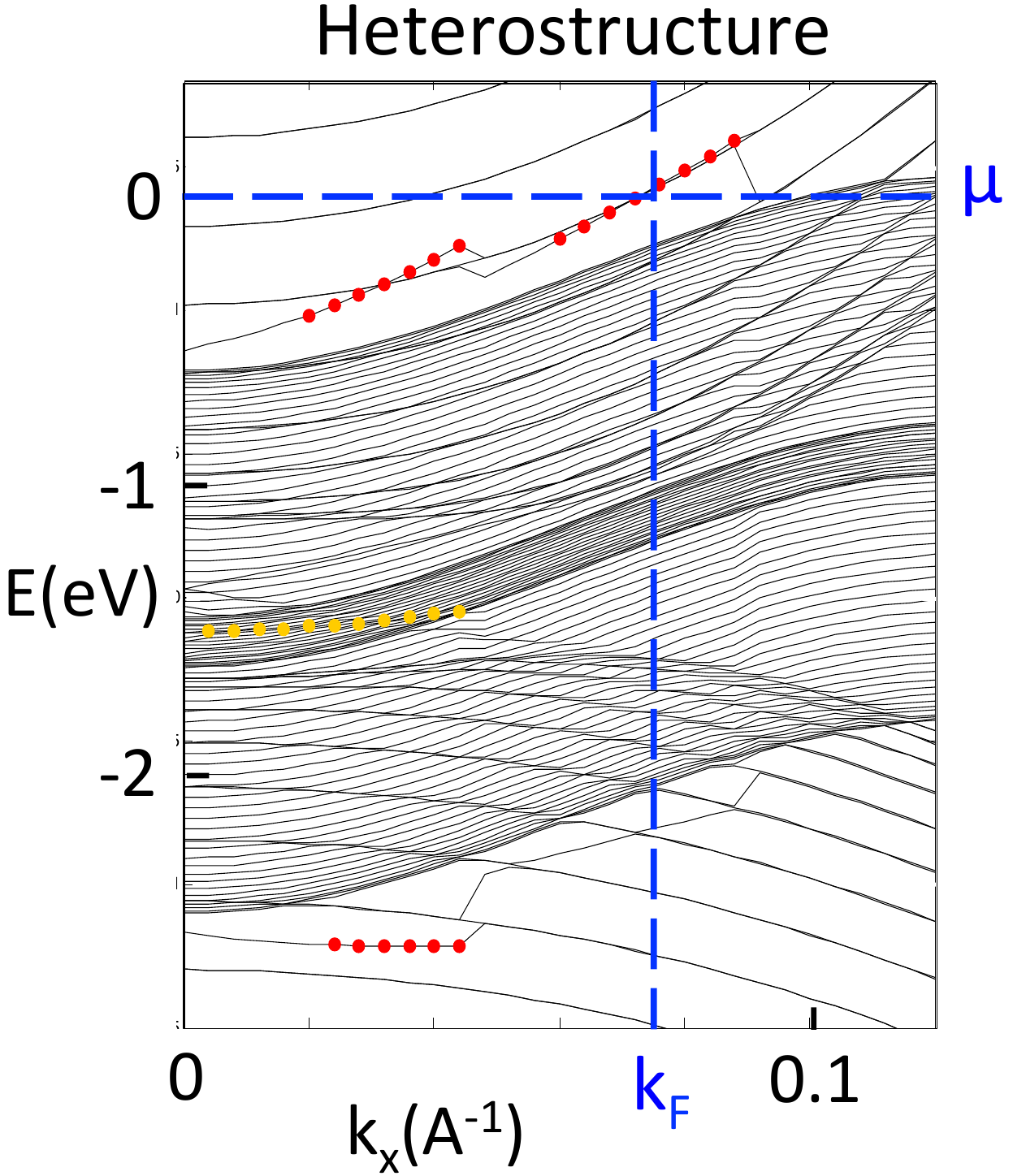}
	}
	\subfigure[]{
	\includegraphics[width=3.5cm]{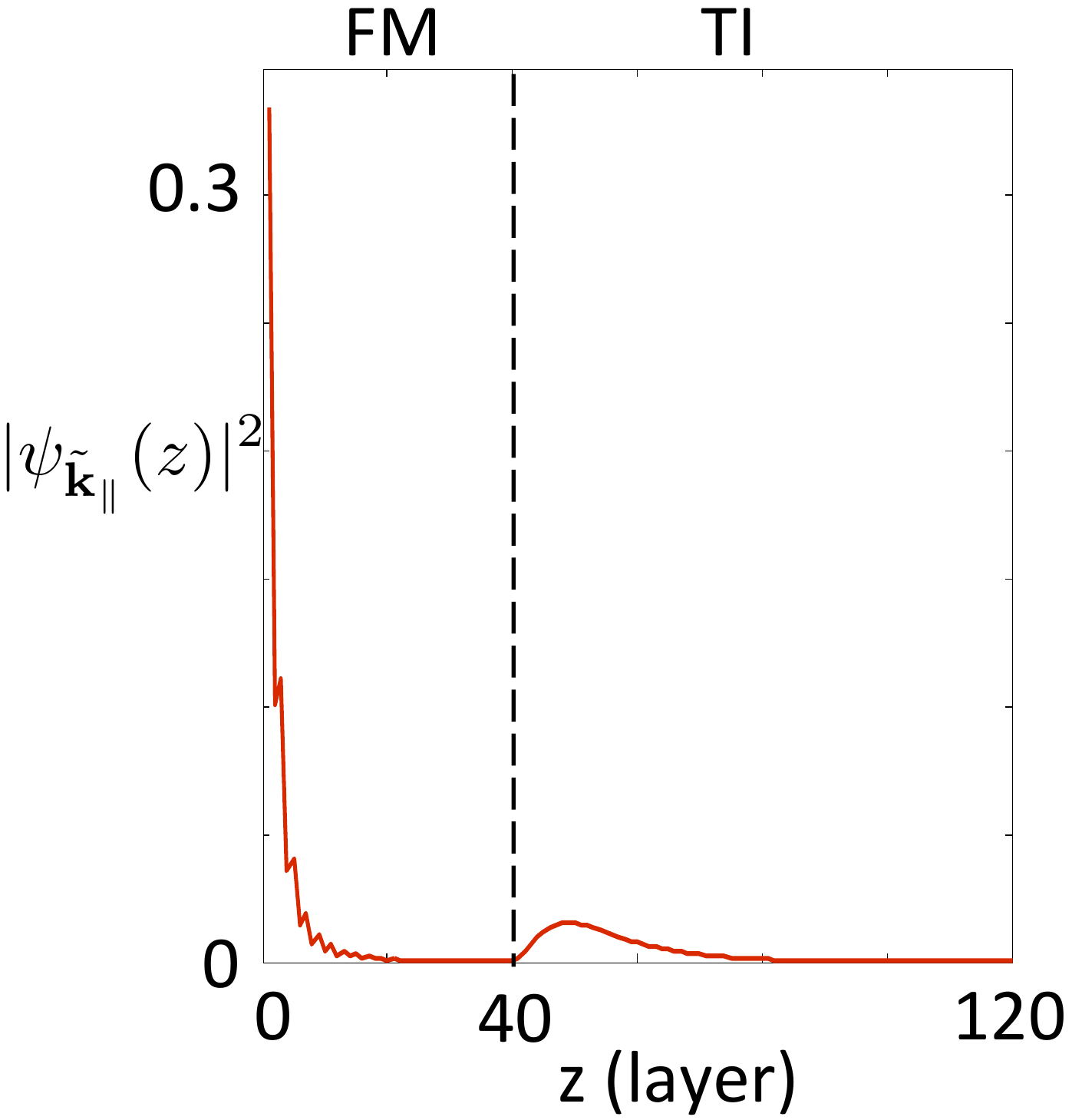}
	}
\caption{
(a) The dispersions of the FM slab (the left panel) and the TI slab (the right panel) before forming heterostructure along in-plane momentum $\mathbf{k}_{\parallel}=(k_x,0)$. The upper and lower dashed lines represent the Fermi levels of the FM ($\mu_{FM}$) and TI slab ($\mu_{TI}$), respectively.
(b) The band structure of the heterostructure in the presence of hybridization. The vertical and horizontal dashed lines represent Fermi level $\mu$ and the Fermi momentum $\tilde{\mathbf{k}}_{\parallel}=(k_F,0)$, respectively. The hybridization strength $g_{\beta}(\mathbf{k_{\parallel}})$ is given in the text with $\tilde{g}_0=0.078$ eV and $\tilde{g}_1=0.098$ eV.
The red and the orange dots label the interface states that emerge outside of and survive within the FM band, respectively. Here, the interface states are defined as states consisting of a substantial weight $>48\%$ on the TI side that has more than $80\%$ of weight localized in the first $30\%$ of the TI slab away from the interface.
(c) The wave function of the interface state in the heterostructure (red dots in (b)) at the Fermi level $\mu$ with momentum ($k_F$,0).
For all subfigures the slab thicknesses are $N_{TI}=80$, $N_m=40$, and the band-bending potential parameters are
$V_0=1$ eV and $\eta=0.3$, respectively.
}
\label{tbspec}
\end{figure}

As for the FM slab,
we model each layer $j$ by a simple two-band model
\begin{equation}
H_{FM}(\mathbf{k_{\parallel}},j)=\mathbbm{m}~c^{\dagger}_{\mathbf{k_{\parallel}},j}c_{\mathbf{k_{\parallel}},j}
+\mathbbm{t}~c^{\dagger}_{\mathbf{k_{\parallel}},j+1}c_{\mathbf{k_{\parallel}},j},
\label{eq:HM}
\end{equation}
where $\mathbbm{m}=(m_{\mathbf{k_{\parallel}},j}-\mu_{FM})\mathbb{1}_{2\times 2}+\Delta\sigma_x$ and $\mathbbm{t}=t_z\mathbb{1}_{2\times 2}$ are $2\times 2$ matrices in spin basis.
Here, $m_{\mathbf{k_{\parallel}},j}=-t_{\parallel}\cos(k_{\parallel}a)$ is the dispersion given by in-plane hopping $t_{\parallel}$ with in-plane lattice constant $a$, $\mu_{FM}$ is the Fermi level, $t_z$ is the hopping in the $z$ direction, and $\Delta$ is the exchange energy where we choose the magnetization to be in the $x$ direction.
The parameters of the FM layer are chosen such that the FM bands overlap with the Dirac interface branch in both  momentum $\mathbf{k_{\parallel}}$ and energy [see Fig.~\ref{tbspec}(a)].

\begin{figure}[t]
	\subfigure[]{
	\includegraphics[width=4cm]{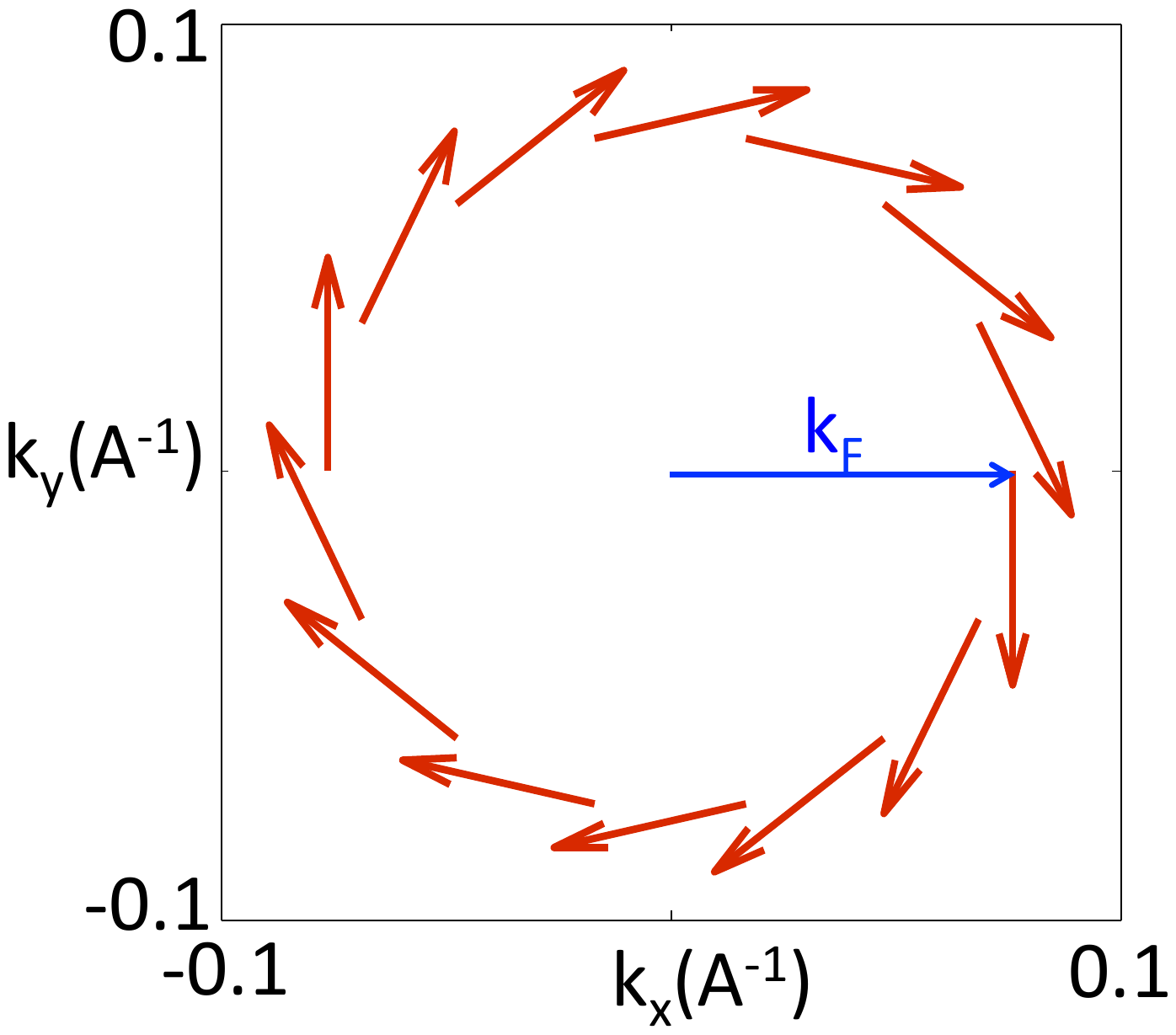}
	}
	\subfigure[]{
	\includegraphics[width=4cm]{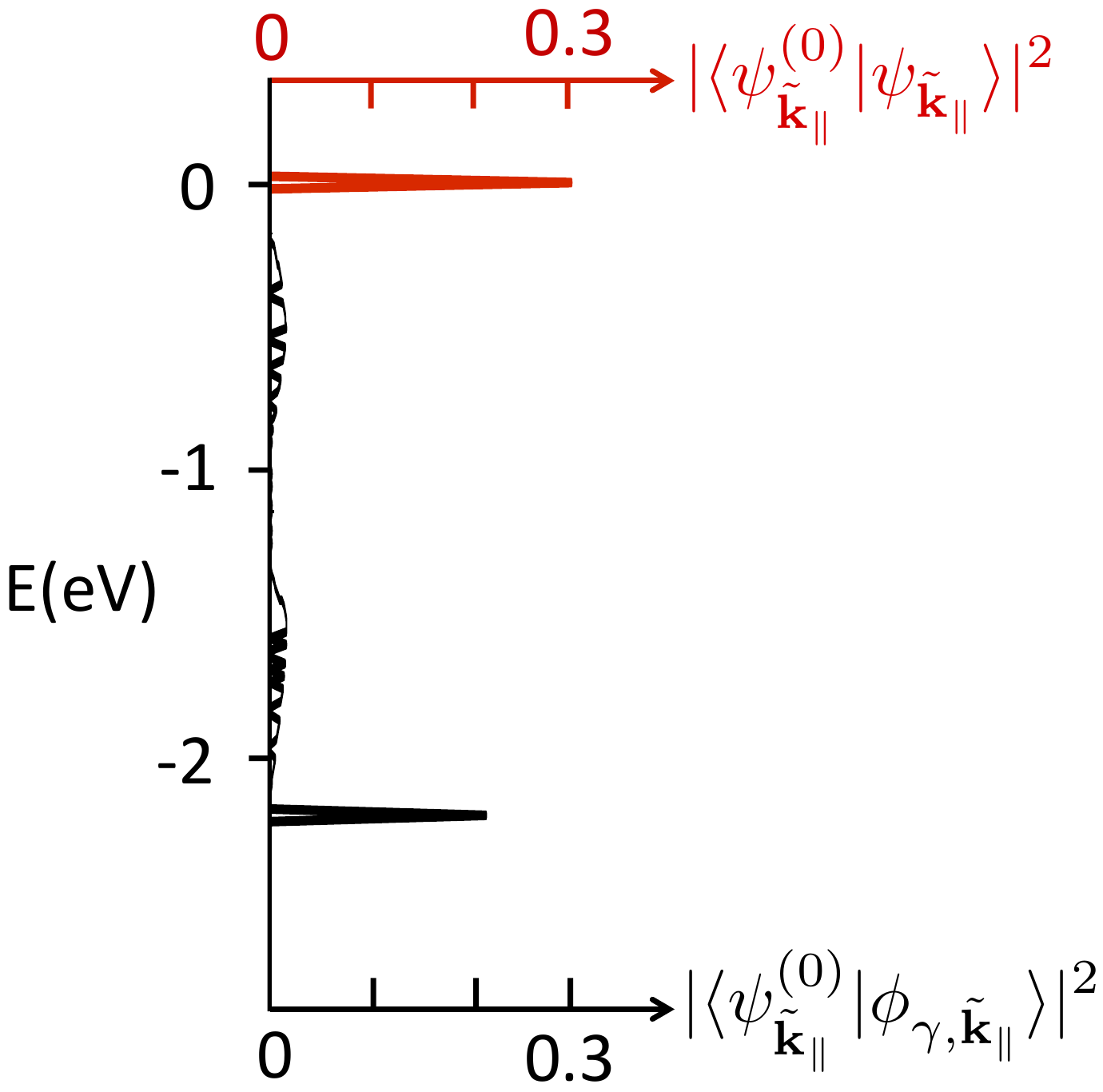}
	}
\caption{(a) The spin texture $(S_{x,\tilde{\mathbf{k}}_{\parallel}},S_{y,\tilde{\mathbf{k}}_{\parallel}})$ of the deescendent state in the hybridized heterostructure at Fermi level $\mu$ with Fermi momenta  $\tilde{\mathbf{k}}_{\parallel}=k_F(\cos\theta_k,\sin\theta_k)$.
(b) The weight of the original Dirac interface state $|\psi^{(0)}_{\tilde{\mathbf{k}}_{\parallel}}\rangle$ carried by the eigenstates of the heterostructure along the momentum-cut at $\tilde{\mathbf{k}}_{\parallel}=(k_F,0)$ (the vertical dashed line in Fig.~\ref{tbspec}(b)). The red (upper) horizontal axis and red curve are for the new interface state $|\psi_{\tilde{\mathbf{k}}_{\parallel}}\rangle$ at the Fermi level $\mu$, and the black (lower) horizontal axis and black curve are for the rest of the eigenstates $|\phi_{\gamma,\tilde{\mathbf{k}}_{\parallel}}\rangle$.
}
\label{tbresult}
\end{figure}

In the absence of hybridization, the bilayer Hamiltonian $H_0(\mathbf{k_{\parallel}})$ can be diagonalized into
\begin{align}
H_{0}(\mathbf{k_{\parallel}})=&(\epsilon^0_{\mathbf{k_{\parallel}}}-\mu_{TI})d_{\mathbf{k_{\parallel}}}^{0\dagger}d^0_{\mathbf{k_{\parallel}}}\nonumber\\
&+\sum_{\alpha=1}^{4N_{TI}-1}(E^0_{TI,\alpha,\mathbf{k_{\parallel}}}-\mu_{TI})b_{\alpha,\mathbf{k_{\parallel}}}^{0\dagger}b^0_{\alpha,\mathbf{k_{\parallel}}}
\nonumber\\
&+\sum_{\beta=1}^{2N_{FM}}(E^0_{FM,\beta,\mathbf{k_{\parallel}}}-\mu_{FM})f_{\beta,\mathbf{k_{\parallel}}}^{0\dagger}f^0_{\beta,\mathbf{k_{\parallel}}},
\label{eq:H0tbdiag}
\end{align}
where the four-spinors $d^0_{\mathbf{k_{\parallel}}}$ and $b^0_{\alpha,\mathbf{k_{\parallel}}}$ annihilate the Dirac interface state with energy $\epsilon^0_{\mathbf{k_{\parallel}}}$ and the rest of the eigenstates in the TI with energy $E^0_{TI,\alpha,\mathbf{k_{\parallel}}}$, respectively, and the two-spinor $f^0_{\beta,\mathbf{k_{\parallel}}}$ annihilates the FM states with energy $E^0_{FM,\beta,\mathbf{k_{\parallel}}}$.
Here, $\alpha$ and $\beta$ label the TI states besides the Dirac interface state and the FM states, respectively.
In Fig.~\ref{tbspec}(a) we plot the spectra of the FM and TI layer before forming a heterostructure; the spectra are given by the dispersion of the model in Eq.~\eqref{eq:H0tbdiag} but with the band-bending potential $V(\mathbf{k_{\parallel}},j)=0$.
The effect of the band-bending potential is to break the degeneracy between the two TI surfaces and shifts the dispersion of the Dirac interface state downwards, which is expected when the localization length is smaller than the well width $1/\eta$\cite{TIInsulThy}.

Now we introduce the hybridization term which preserves in-plane momenta $\mathbf{k_{\parallel}}$ and spin:
\begin{equation}
H'(\mathbf{k_{\parallel}})=\sum_{\beta=1}^{2N_{FM}}g_{\beta}(\mathbf{k_{\parallel}})\; f_{\beta,\mathbf{k_{\parallel}}}^{0\dagger}d^0_{\mathbf{k_{\parallel}}}+H.c.,
\label{eq:hhyb}
\end{equation}
where $\beta$ runs over all the FM states at $\mathbf{k_{\parallel}}$.
Here, we expect the strength $g_{\beta}(\mathbf{k_{\parallel}})$ to be proportional to the `spin-overlap' between the FM states and the spin-momentum-locked Dirac interface state:
\begin{equation}
g_{\beta}(\mathbf{k_{\parallel}})=\tilde{g}_{\beta}(\mathbf{k_{\parallel}})\langle\Psi^{0}_{\mathbf{k_{\parallel}}}|\Phi^0_{\beta,\mathbf{k_{\parallel}}}\rangle,
\label{eq:gspindep}
\end{equation}
where $|\Phi^0_{\beta,\mathbf{k_{\parallel}}}\rangle\equiv\sum_{z'}\langle z'|\phi^0_{\beta,\mathbf{k_{\parallel}}}\rangle$ and $|\Psi^0_{\mathbf{k_{\parallel}}}\rangle\equiv\sum_{z,a}\langle z,a|\psi^0_{\mathbf{k_{\parallel}}}\rangle$ are `coarse-grained' two-component ket spinors in spin basis for the FM and unhybridized Dirac interface states, respectively.
Here we define $|\phi^{0}_{\beta,\mathbf{k_{\parallel}}}\rangle\equiv f_{\beta,\mathbf{k_{\parallel}}}^{0\dagger}|0\rangle$, $|\psi^{0}_{\mathbf{k_{\parallel}}}\rangle\equiv d_{\mathbf{k_{\parallel}}}^{0\dagger}|0\rangle$, $z^{(')}$ runs over the TI (FM) layers, and $a=P_1, P_2$ labels the TI atomic orbitals.
For simplicity, we assume $\tilde{g}_{\beta}(\mathbf{k_{\parallel}})=\tilde{g}(\mathbf{k_{\parallel}})$ to be identical for all FM states $\beta$.
To mimic our first-principle-calculated band structure later shown in Fig.~\ref{fig:DFTbands}(a), we further assume $\tilde{g}(\mathbf{k_{\parallel}})$ to increase with the momentum and take $\tilde{g}_{\mathbf{k_{\parallel}}}=\tilde{g}_0+\tilde{g}_1k_{\parallel}$ with $\tilde{g}_0,\tilde{g}_1>0$.
Finally, the full Hamiltonian including hybridization reads $H=\sum_{\mathbf{k}_{\parallel}}H(\mathbf{k_{\parallel}})$ where
\begin{align}
H(\mathbf{k_{\parallel}})=&H_0(\mathbf{k_{\parallel}})+H'(\mathbf{k_{\parallel}})
\label{eq:Htot}
\end{align}
is given by Eq.~\eqref{eq:H0tbdiag} and \eqref{eq:hhyb}.

\section{The descendent state at the interface from the lattice model}

The dispersion of $H$ in Eq.~\eqref{eq:Htot} shows new interface states  [red dots in Fig.~\ref{tbspec}(b)] formed above and below the FM bands as well as the ``remnant states'' [orange dots in Fig.~\ref{tbspec}(b)] which are residues of the Dirac interface states within the FM bands.
Since the upper new interface branch emerges right above the upper band edge of FM, it is likely to intersect the Fermi level $\mu$ of the heterostructure, which is approximately given by that of the FM slab $\mu_{FM}$.
We will thus focus only on this upper new interface branch for the rest of the article.
To examine the properties of these states, we write the diagonalized full Hamiltonian at each $\mathbf{k_{\parallel}}$ as
\begin{align}
H(\mathbf{k_{\parallel}})
&=(\epsilon_{\mathbf{k_{\parallel}}}-\mu)d_{\mathbf{k_{\parallel}}}^{\dagger}d_{\mathbf{k_{\parallel}}}\nonumber\\
&+\sum_{\gamma=1}^{4N_{TI}+2N_{FM}-1}(E_{\gamma,\mathbf{k_{\parallel}}}-\mu)b_{\gamma,\mathbf{k_{\parallel}}}^{\dagger}b_{\gamma,\mathbf{k_{\parallel}}},
\label{eq:Htotdiag}
\end{align}
where $d_{\mathbf{k_{\parallel}}}$ annihilates the new interface state with energy $\epsilon_{\mathbf{k_{\parallel}}}$ above the FM upper band edge, and $b_{\gamma,\mathbf{k_{\parallel}}}$ annihilate the rest of the eigenstates labeled by $\gamma$ with energy $E_{\gamma,\mathbf{k_{\parallel}}}$.
Here  we define
$|\psi_{\mathbf{k_{\parallel}}}\rangle\equiv d_{\mathbf{k_{\parallel}}}^{\dagger}|0\rangle$, $|\phi_{\gamma,\mathbf{k_{\parallel}}}\rangle\equiv b_{\gamma,\mathbf{k_{\parallel}}}^{\dagger}|0\rangle$.
The spatial profile of the upper new interface state is then given by
$|\psi_{\mathbf{k}_{\parallel}}(z)|^2$, which is a function of $z$ measured from the bottom of the FM slab along the finite dimension of the heterostructure.
$|\psi_{\mathbf{k}_{\parallel}}(z)|^2$ at the Fermi momentum $\tilde{\mathbf{k}}_{\parallel}=(k_F,0)$ [see Fig.~\ref{tbspec}(c)] shows that the TI portion of the upper new interface at the Fermi level localizes near the interface.

The new interface state at the Fermi level, i.e. the descendent state, also has a clockwise Rashba-type spin-winding just like the origianl Dirac interface state, as shown by the in-plane spin expectation values $(S_{x,\tilde{\mathbf{k}}_{\parallel}},S_{y,\tilde{\mathbf{k}}_{\parallel}})$ in Fig.~\ref{tbresult}(a). Here, $S_{x/y,\tilde{\mathbf{k}}_{\parallel}}\equiv\langle\psi_{\tilde{\mathbf{k}}_{\parallel}}|\hat{S}_{x/y}|\psi_{\tilde{\mathbf{k}}_{\parallel}}\rangle$ where $\hat{S}_{x/y}\equiv(\mathbb{0}_{2\times2}\otimes\mathbb{0}_{N_{FM}\times N_{FM}})\oplus(\sigma_{x/y}\otimes\mathbb{1}_{2\times 2}\otimes\mathbb{1}_{N_{TI}\times N_{TI}})$ are the spin operators projected onto the TI side.
The spin magnitude of states with negative $k_y$ is slightly smaller than that with positive $k_y$ due to the spin-dependence of hybridization strength in Eq.~\eqref{eq:gspindep}, which vanishes in the limit where the ferromagnetic exchange energy $\Delta$ vanishes.

Finally, to determine the origin of this new interface state $|\psi_{\mathbf{k_{\parallel}}}\rangle$, we examine the weight of the original Dirac interface state $|\psi^{0}_{\mathbf{k_{\parallel}}}\rangle$ carried by the heterostructure eigenstates $|\psi_{\mathbf{k}_{\parallel}}\rangle$ and $|\phi_{\gamma,\mathbf{k}_{\parallel}}\rangle$.
Fig.~\ref{tbresult}(b) shows the spectral distribution of the weight $\langle\psi^{0}_{\tilde{\mathbf{k}}_{\parallel}}|\psi_{\tilde{\mathbf{k}}_{\parallel}}\rangle$ and $\langle\psi^{0}_{\tilde{\mathbf{k}}_{\parallel}}|\phi_{\gamma,\tilde{\mathbf{k}}_{\parallel}}\rangle$ along the momentum-cut at $\tilde{\mathbf{k}}_{\parallel}=(k_F,0)$ where
two peaks form outside of the FM band, which resembles that of the generic Fano model\cite{Mahan}.
Specifically, the stronger peak is contributed by the new interface state on the Fermi level $|\psi_{\tilde{\mathbf{k}}_{\parallel}}\rangle$ [see the red curve] while the weaker peak buried below the Fermi level $\mu$ and the residues within the band are contributed by the rest of the states $|\phi_{\gamma,\tilde{\mathbf{k}}_{\parallel}}\rangle$ [see the black curve].
Hence we have demonstrated that the hybridization-induced new interface state at the Fermi level is the descendent state of the original Dirac interface state, which thus inherits the wavefunction localization and spin texture of the mother state.

\section{First-principles-based simulation of TI-FM bilayer}

\begin{figure}[h]
\includegraphics[width=0.2\textwidth]{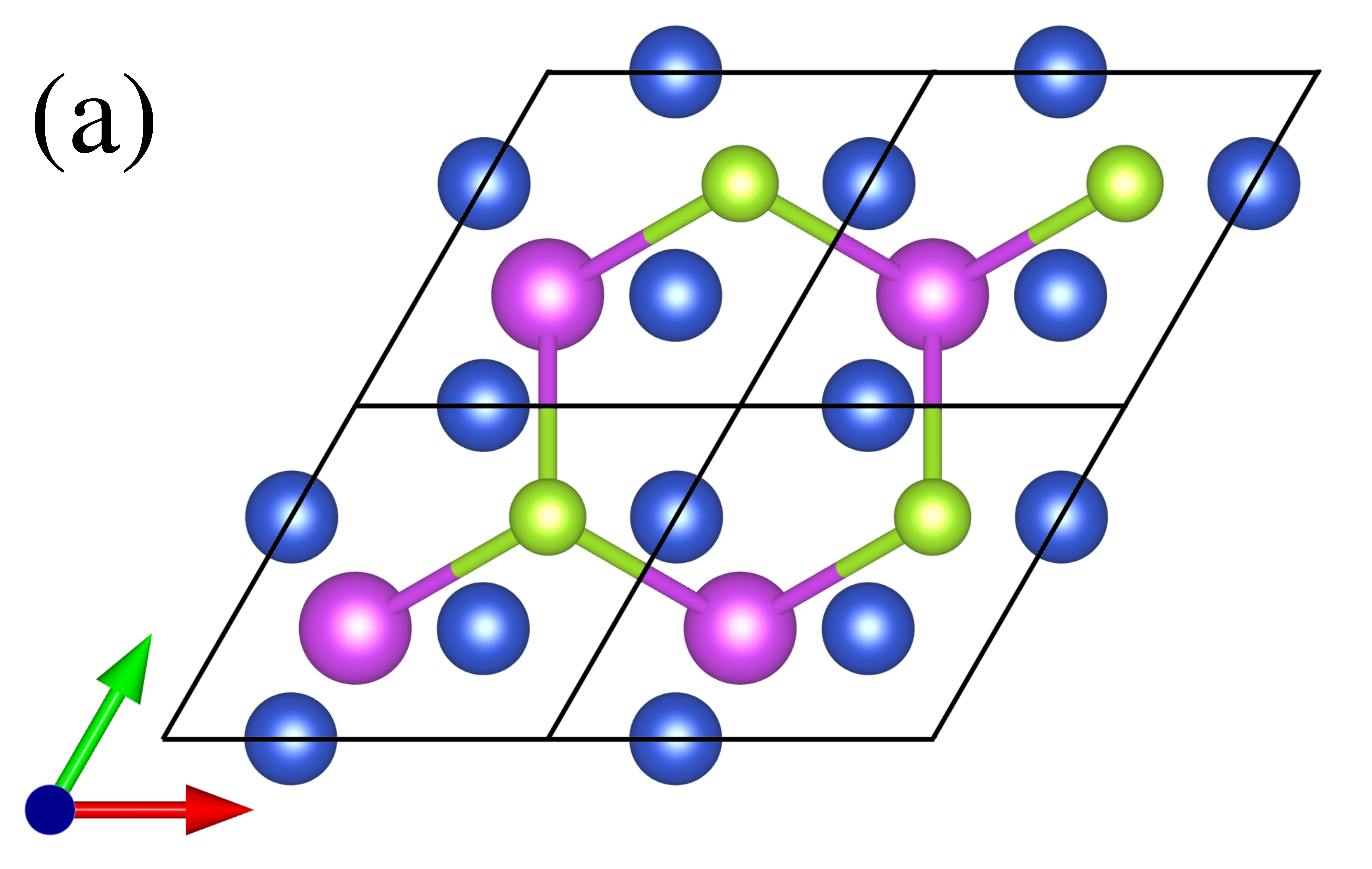}
\hspace{0.6truecm}
\includegraphics[width=0.2\textwidth]{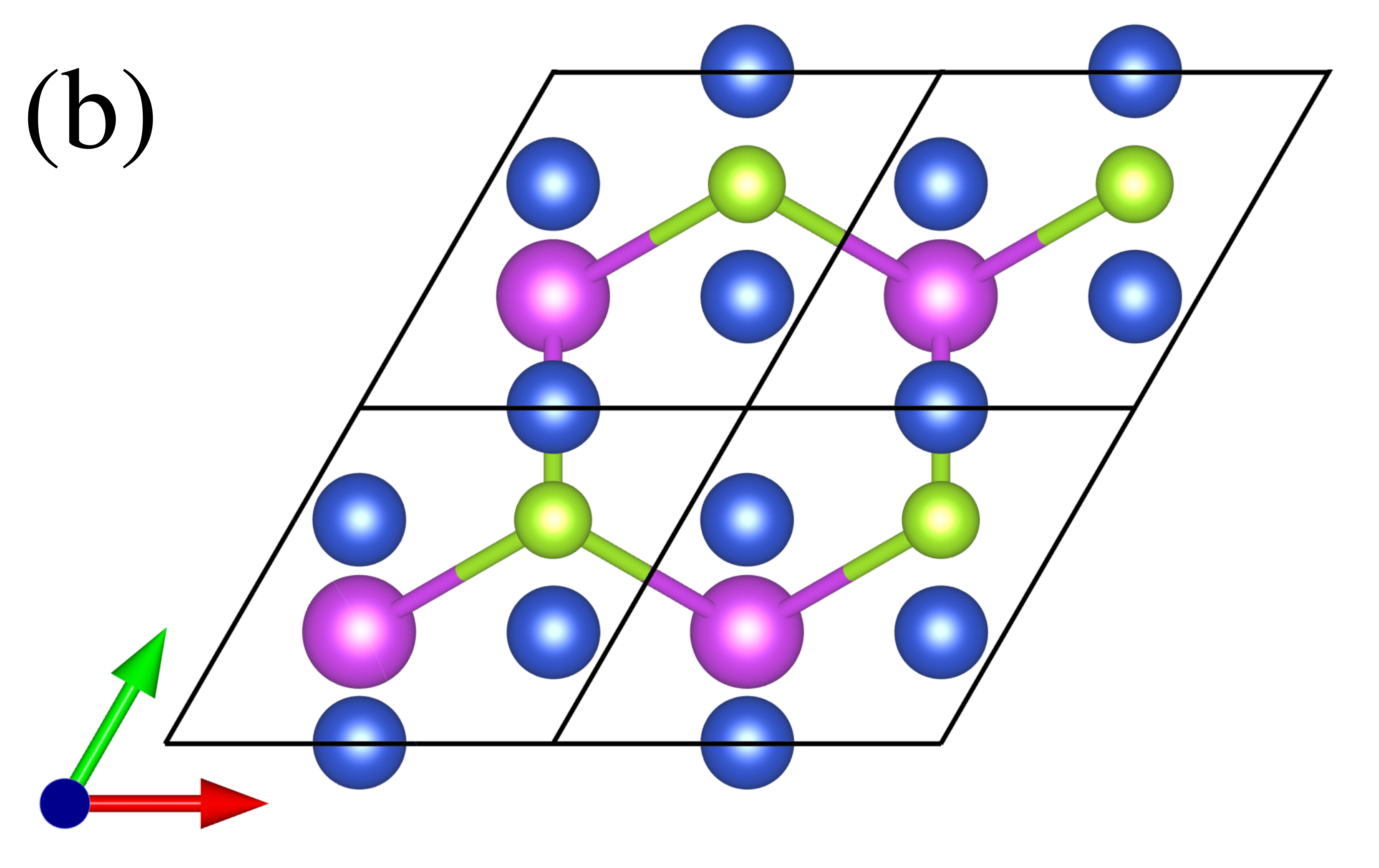}
\caption{Top views of two different interfaces for the TI-FM bilayer. Interface (a) has a lower energy than interface (b) by
0.07~eV upon the geometry relaxation. Only topmost Se (small green), Bi (large purple), and Ni (medium blue) atoms at the interface are shown. The in-plane lattice vectors in real space are shown as red and green arrows.}
\label{fig:interfaces}
\end{figure}

We simulate a TI-FM bilayer within DFT by using {\tt VASP} \cite{VASP}. We use the generalized gradient approximation (GGA) \cite{PERD1996} for the exchange-correlation functional and projector-augmented wave (PAW) pseudopotentials \cite{PAW,VASP}.
Spin-orbit coupling (SOC) is included self-consistently with the DFT calculation. We construct a TI-FM bilayer by using a supercell consisting
of a $1 \times 1 \times$ 5-QL slab of Bi$_2$Se$_3$(111) beneath a Ni(111) slab of $\sqrt{3} \times \sqrt{3} \times L_z$, where $L_z$ is
4 atomic layers, with a thick vacuum layer of 35.7~\AA.~The in-plane lattice constant of the supercell is fixed as the experimental lattice constant of the TI, 4.143~\AA~\cite{NAKA63}. This gives rise to a 4\% compressive strain onto the Ni slab. We consider two different interface geometries shown
in Fig.~\ref{fig:interfaces}. The $x$, $y$, and $z$ coordinates of the QL nearest to the interface and the $x$, $y$, and $z$ coordinates of all the Ni atoms in the supercell are relaxed until the residual forces are lower than 0.1~eV/\AA,~while keeping the atomic coordinates of the rest of the QLs in the TI remain fixed. Interface (a) (considered in Ref.~\onlinecite{ThyTIFM}) gives a lower energy than interface (b) by 0.07~eV upon the relaxation. Therefore, henceforth, we present results obtained from interface (a). The distance between the TI and Ni layer is found to be 1.98~\AA~after the relaxation, which agrees with the previous DFT calculations of the TI-FM bilayer \cite{TImetalDFT,ThyTIFM}. For the TI-FM bilayer, $11 \times 11 \times 1$ $k$ points are sampled in the geometry relaxation and calculations of electronic structure.
The $z$ axis is along the [111] direction and the $y$ axis is along the [11$\bar{2}$] direction in the TI rhombohedral structure. In our
simulation the magnetization of the Ni slab is in plane, such as parallel to the $y$ axis.

\section{The descendent state from the simulation}

\begin{figure}[h]
\includegraphics[width=0.23\textwidth]{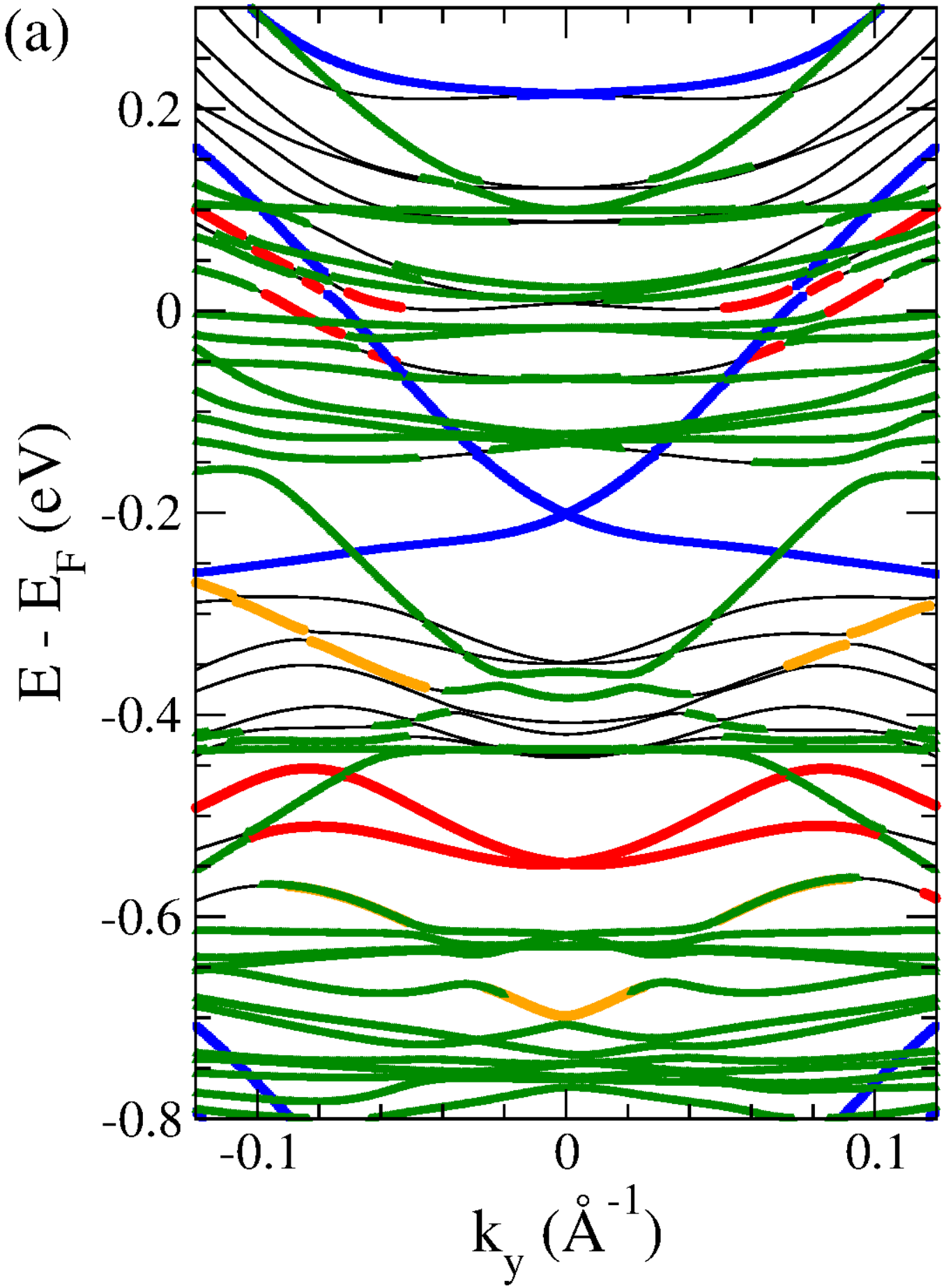}
\hspace{0.6truecm}
\includegraphics[width=0.19\textwidth]{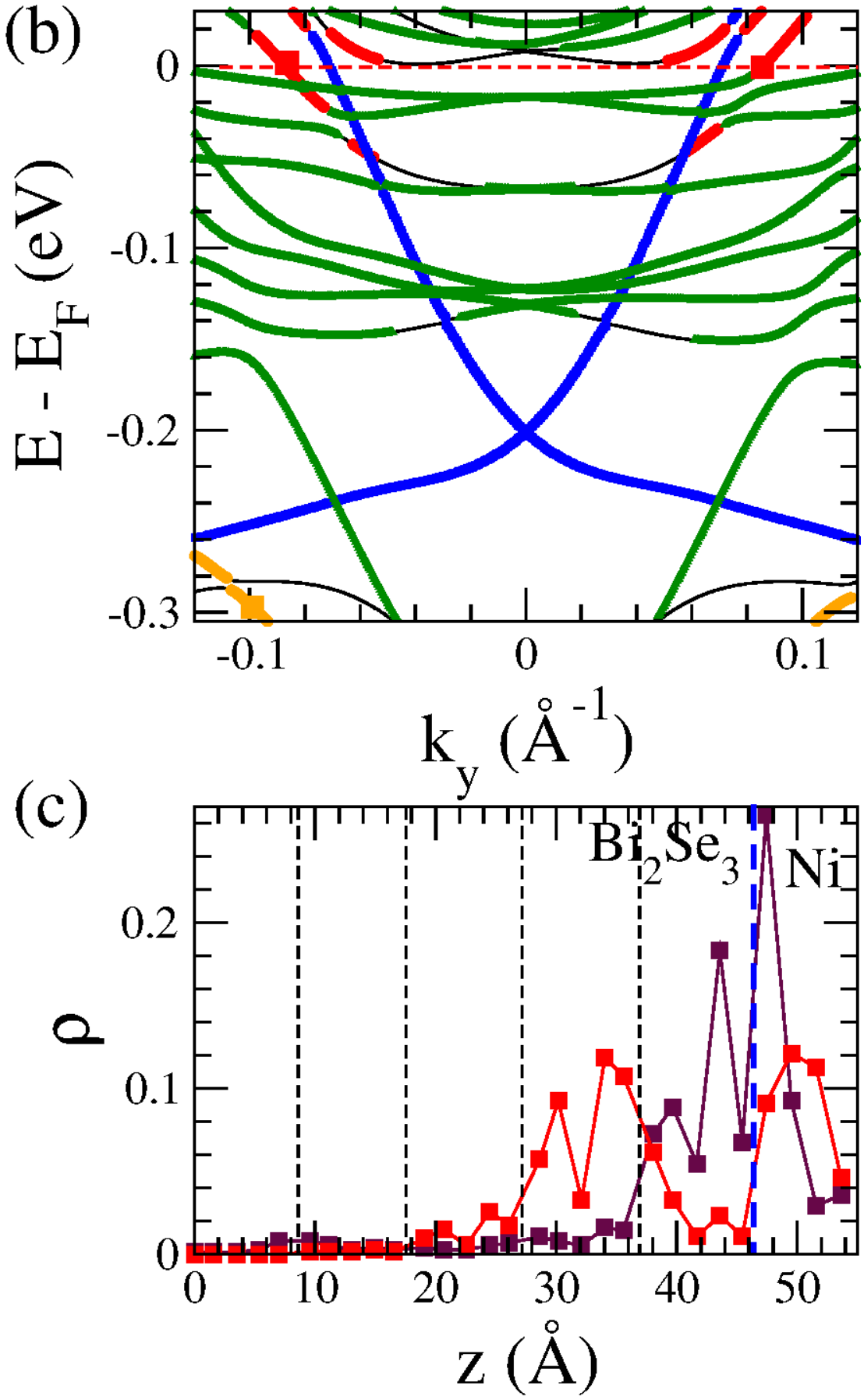}
\caption{(a) DFT-calculated band structure of the TI-FM bilayer along the $k_y$ axis and (b) zoom-in of (a) when the magnetization is along the $y$ axis. 
In (a) and (b), green (blue) symbols represent states localized into the Ni slab (the bottommost QL in contact to vacuum), while red symbols are for states 
localized into the topmost-1 QL, like descendent states. Orange symbols are for the states localized at the topmost QL.
(c) Relative electron density $\rho(z)$ profile (dimensionless) of the red and orange bands as a function of $z$ coordinates. The red color is for the $\rho(z)$ value 
of the red band or decendent state computed at $k_y=-0.087$~\AA$^{-1}$ or $k_{\mathrm F}$ (red squares in (b)), while the maroon color is for the $\rho(z)$ value of the orange band 
at $k_y=-0.1$~\AA$^{-1}$ (orange square near $-0.3$~eV in (b)). }
\label{fig:DFTbands}
\end{figure}

\begin{figure}[h]
\includegraphics[width=0.21\textwidth]{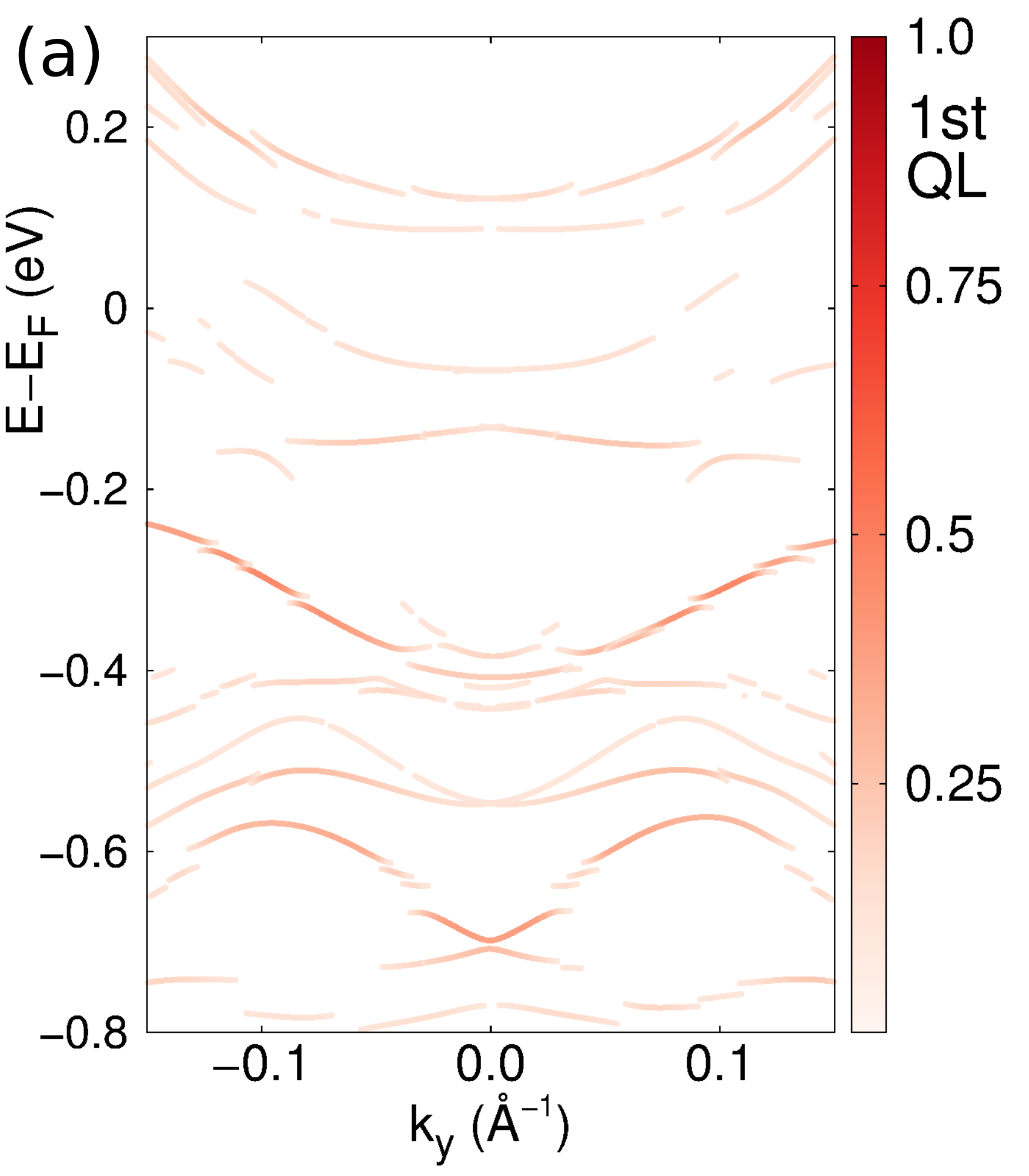}
\hspace{0.3truecm}
\includegraphics[width=0.21\textwidth]{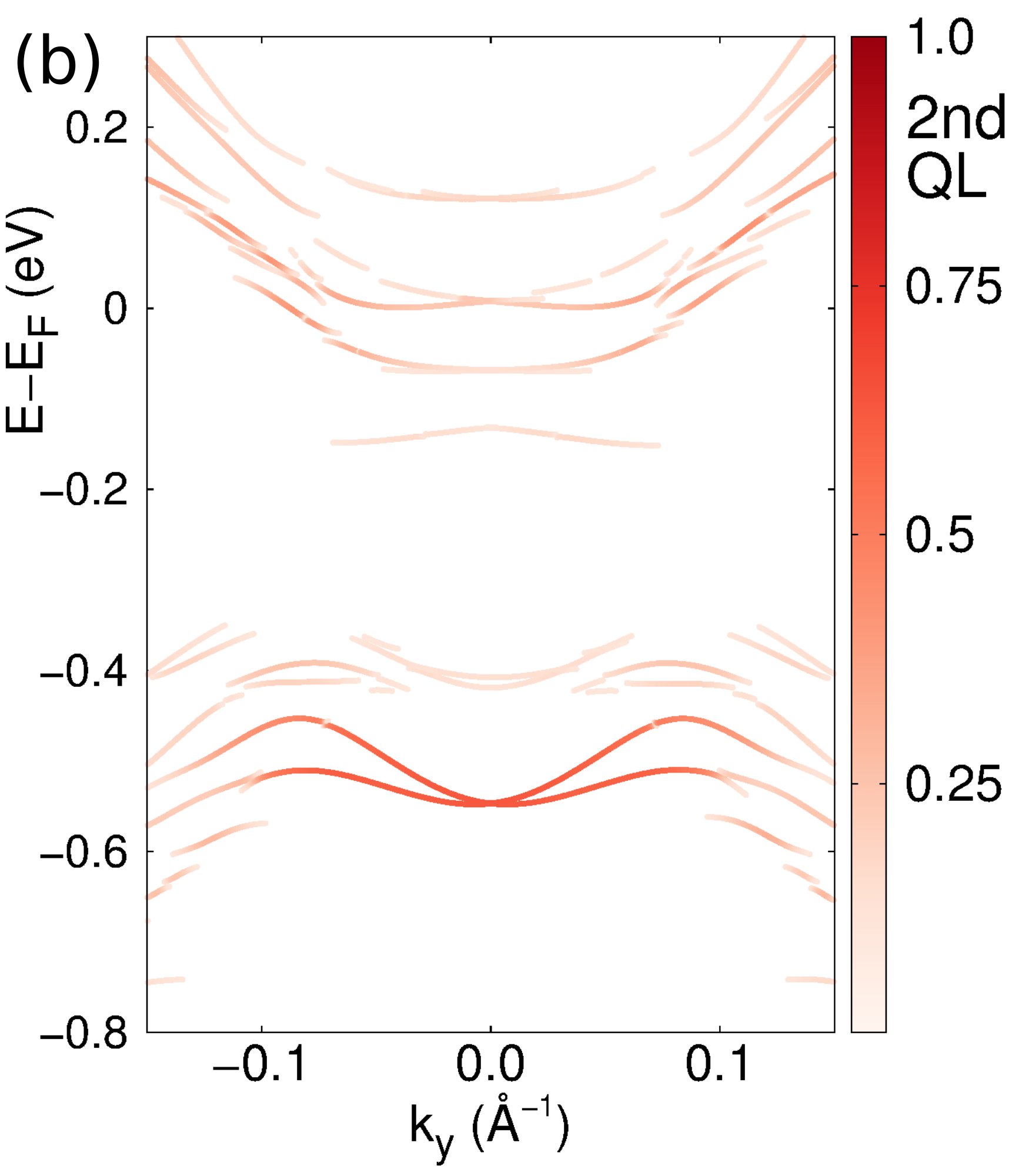}
\caption{DFT-calculated band structure of the TI-FM bilayer, same as Fig.~\ref{fig:DFTbands}(a), but showing only the states (a) with the weight of the topmost QL
greater than 10\%, and (b) with the weight of the topmost-1 QL greater than 10\%. The color box scale is the percentage weight of the topmost QL for (a) and 
at the topmost-1 QL for (b).}
\label{fig:DFTadd}
\end{figure}

\begin{figure}[h]
\includegraphics[width=0.21\textwidth]{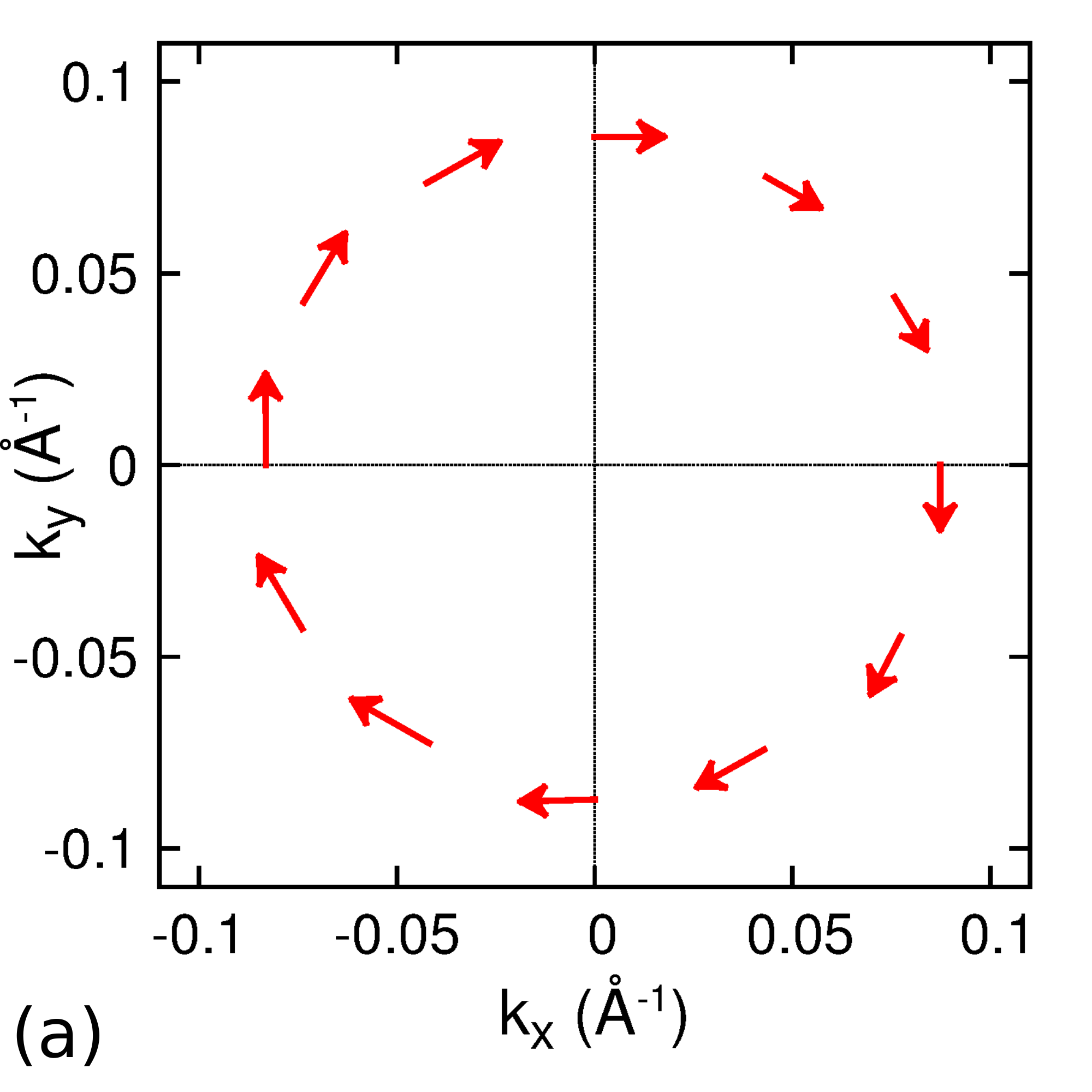}
\hspace{0.5truecm}
\includegraphics[width=0.21\textwidth]{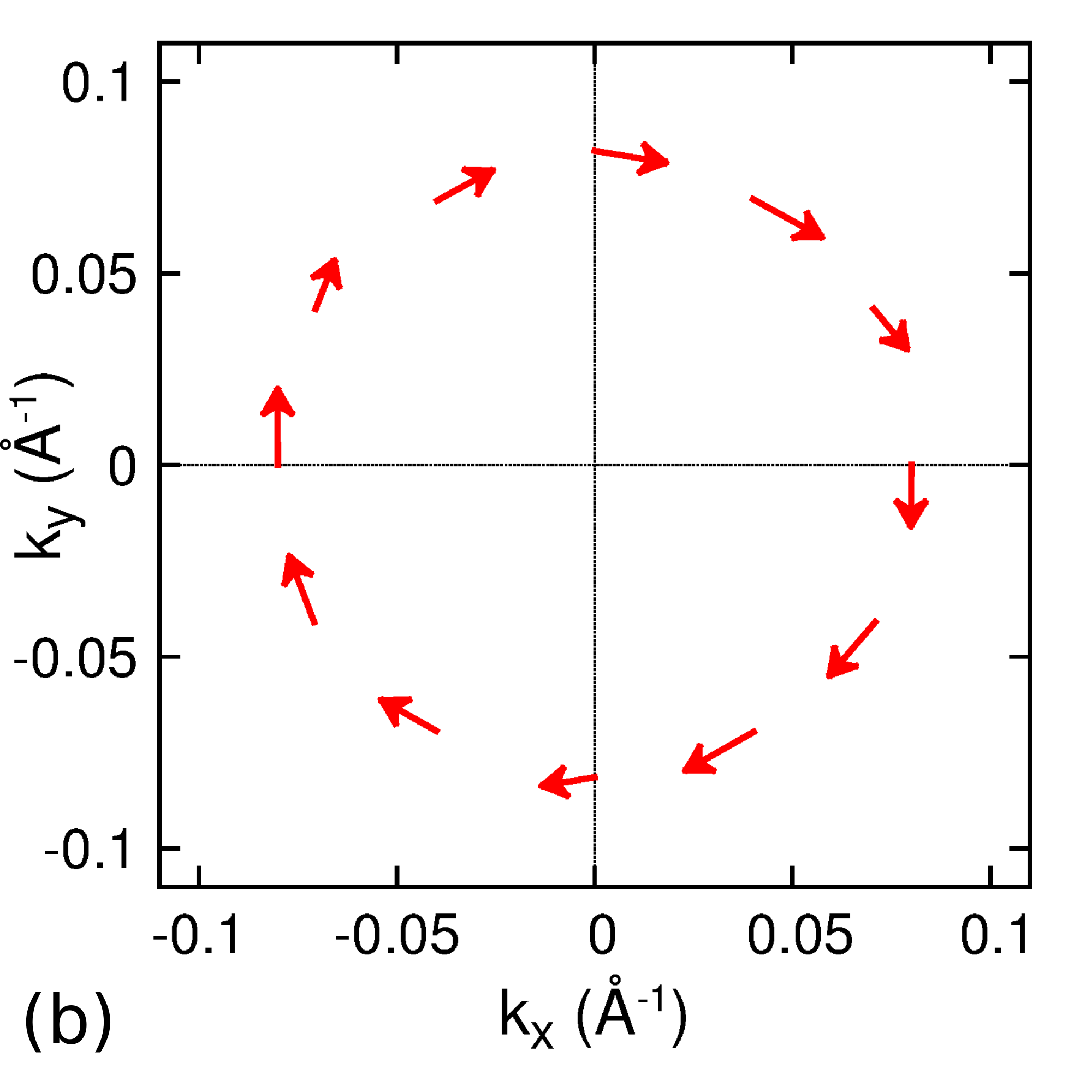}
\caption{(Color online) (a) DFT-calculated spin texture of the descendent state projected onto the TI side at $E_F$ indicated as the red dashed line in 
Fig.~\ref{fig:DFTbands}(b) when the magnetization is along the $y$ axis. This spin texture can be compared to Fig.~\ref{tbresult}(a). 
(b) DFT-calculated spin texture of the descendent state projected onto the TI side at $E_F$ when the magnetization is along the $z$ axis.}
\label{fig:DFTTIspin}
\end{figure}

\begin{figure}[h]
\includegraphics[width=0.21\textwidth]{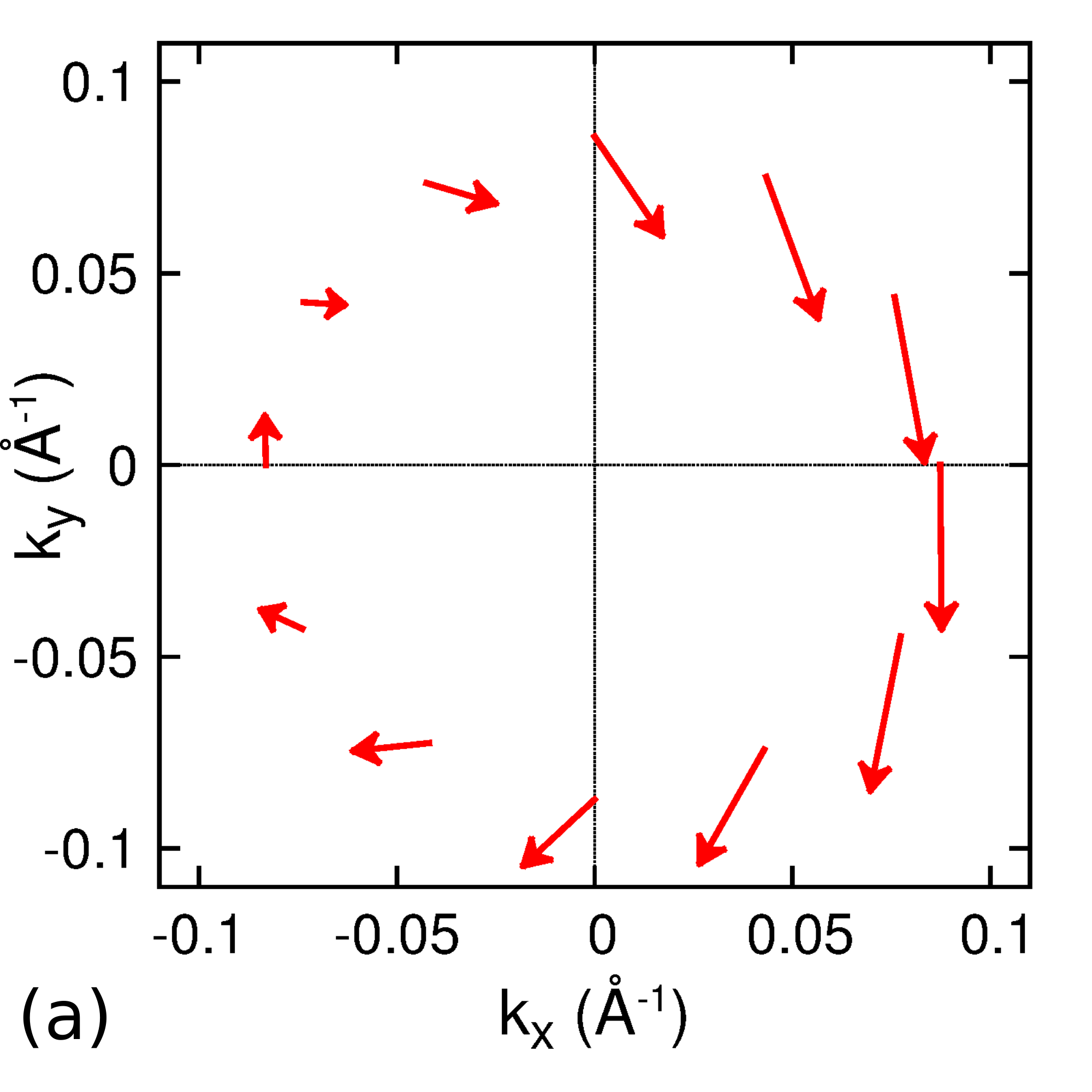}
\hspace{0.5truecm}
\includegraphics[width=0.21\textwidth]{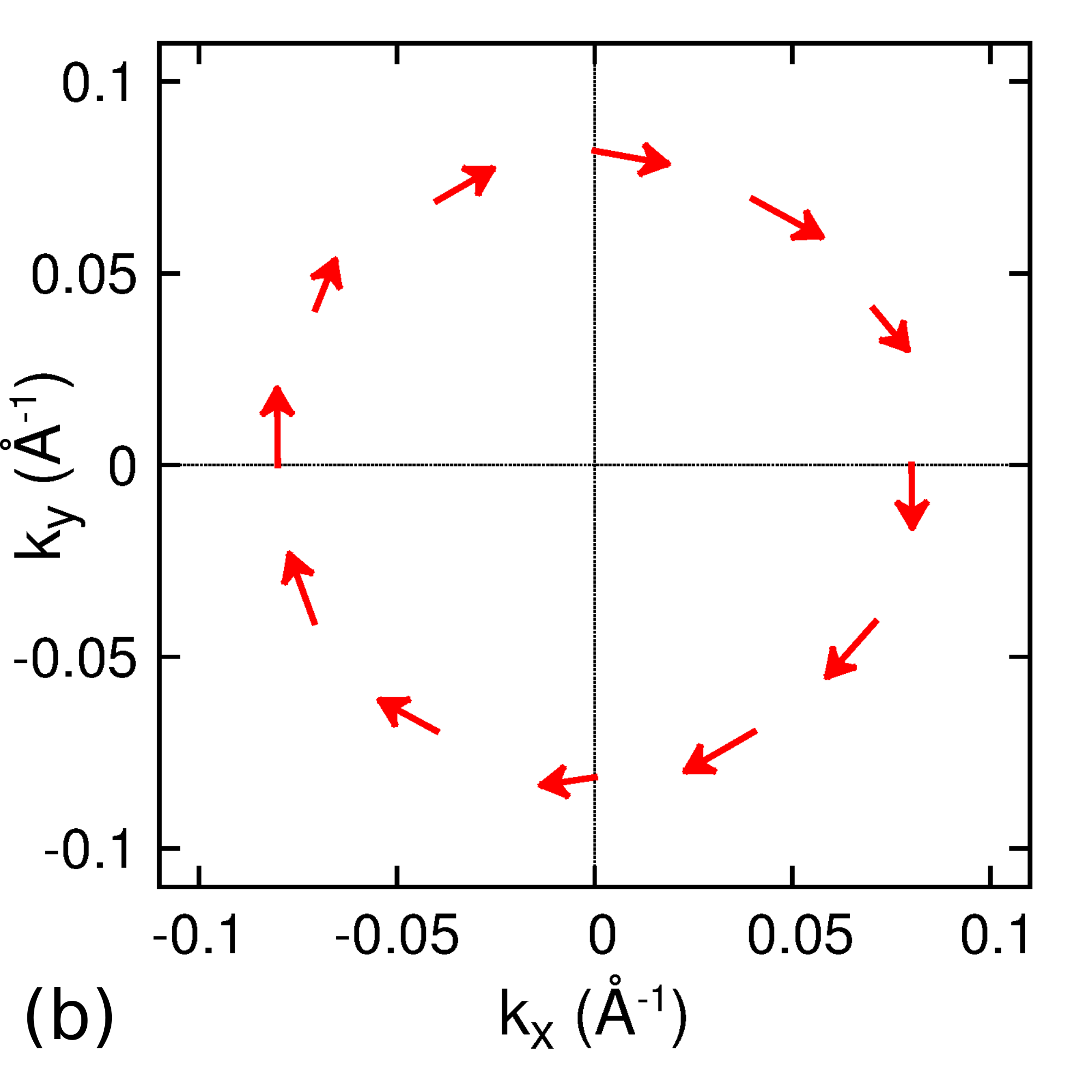}
\caption{(Color online) DFT-calculated spin texture of the descendent state at $E_F$ when the magnetization is along (a) the $y$ axis and (b) the $z$ axis.
Here the contributions of the Ni slab and the TI film are all included.}
\label{fig:DFTspin}
\end{figure}

We calculate band structure of the TI-FM bilayer along different directions in the $k_x-k_y$ plane such as $\theta_k$=0, $\pm \pi/6$,
$\pm \pi/3$, $\pm 2\pi/3$, $\pm 5\pi/6$, $\pm \pi/2$, and $\pi$, where $\theta_k$ is the azimuthal angle in the $k_x-k_y$ plane.
The band structure along the $k_y$ axis is shown in Fig.~\ref{fig:DFTbands}(a). In the vicinity of $E_F$ the Dirac cone localized at the QL
closest to the interface (topmost QL) is not found, although many bands from the Ni slab [green bands in Fig.~\ref{fig:DFTbands}(a)] appear.
Our DFT+SOC calculations show some charge transfer from the Ni slab to the TI slab, which is caused by the difference between the chemical 
potentials of the TI and Ni slabs (the work functions of the TI and Ni slabs are 5.51 and 5.27~eV, respectively). 
This charge transfer shifts the Dirac surface states at the interface downward far below $E_F$ and causes
very strong hybridization with the Ni states. In addition, the strong hybridization also induces significant relaxations in the $z$ coordinates
of the topmost QL due to the Ni slab. These two factors make it difficult to discern the original Dirac interface state in the vicinity of $E_F$,
which is consistent with the result of the microscopic lattice model discussed in Sec.  III.

There are two classes of states near $E_F$. One class of states [orange bands in Fig.~\ref{fig:DFTbands}(a)] are localized at the topmost QL 
[maroon curve in Fig.~\ref{fig:DFTbands}(c)] with energies in the range of $-0.37 \sim -0.23$~eV somewhat away from the $\Gamma$ point (and 
in the vicinity of $-0.7$~eV near the $\Gamma$ point). These states correspond to the ``remnant states'' found in the microscopic lattice model. 
The other more prominent class of states (red bands in Fig.~\ref{fig:DFTbands}(a) and (b)) are mostly localized into the topmost-1 QL [red curve
in Fig.~\ref{fig:DFTbands}(c)]. These states resemble the descendent states of the lattice model in that (i) they appear near $E_F$, (ii) 
they do not exhibit Dirac-like dispersion, and (iii) their wave function distribution is similar to that from the lattice model.
Figure~\ref{fig:DFTadd}(a) and (b) show the same DFT-calculated band structure with the weight of the topmost QL and of the topmost-1 QL marked 
in color boxes, respectively.

In addition, we examine the spin texture of the descendent state at $E_F$ by computing expectation values of the $x$ and $y$ components of spin, 
$\langle S_{x,\tilde{\mathbf{k}}_{\parallel}} \rangle$ and $\langle S_{y,\tilde{\mathbf{k}}_{\parallel}} \rangle$ at the aforementioned different $\theta_k$ values, 
within DFT. Figure~\ref{fig:DFTTIspin}(a) shows the DFT-calculated spin texture projected onto the TI layer, whereas Fig.~\ref{fig:DFTspin}(a)
reveals the total spin texture including the contribution of the FM layer. As shown in Fig.~\ref{fig:DFTTIspin}(a), 
the DFT-calculated TI spin texture shows two main features: (i) spin-momentum locking; (ii) the magnitude of the spin polarization depending 
on in-plane momentum. The second feature is due to the in-plane magnetization of the Ni slab. 
Our DFT-calculated spin texture is comparable to that from the microscopic model [Fig.~\ref{tbresult}(a)], considering that
the in-plane magnetization of the Ni side in the tight-binding model is along the $x$ axis. Now the DFT-calculated total spin texture
of the descendent state, Fig.~\ref{fig:DFTspin}(a), also shares the similar main features to the projected spin texture with a much stronger 
dependence on in-plane momentum, due to the Ni magnetization. We find that the contribution of the Ni slab to the descendent state varies 
from 26\% to 57\% with the in-plane momentum, which leads to the in-plane momentum dependent coupling to the Ni slab.

We briefly discuss comparison of our DFT calculation to the previous DFT calculations of a TI-Ni bilayer \cite{ThyTIFM,TImetalDFT} and
TI-Co bilayer \cite{ThyTIFM,TEJA2017}. In these previous studies the spin-momentum locking we found above was either not 
reported \cite{ThyTIFM,TImetalDFT} or found well below $E_F$ \cite{TEJA2017}. There are four main differences between ours and the previous 
DFT calculations in addition to a different FM layer in Refs.~\onlinecite{ThyTIFM,TEJA2017}: (i) different magnetization direction of the FM layer, 
(ii) different spatial localization of the state investigated for the spin-momentum locking, (iii) different TI layer thickness, and (iv) supercell
geometry without vacuum vs slab geometry. Among the four differences, the second, third, and fourth are crucial factors. We propose that the descendent 
state (localized at topmost-1 QL) near $E_F$ contributes to the large enhancement of the spin-transfer torque in the TI-FM bilayer in experiment\cite{DanExp}, 
while the previous studies were searching for the states localized at the topmost QL \cite{ThyTIFM,TImetalDFT,TEJA2017}.  
In Ref.~\onlinecite{ThyTIFM}, the TI layer considered, i.e. 3-QL of pristine Bi$_2$Se$_3$, is too thin to form a gapless TI Dirac cone, as experimentally 
confirmed \cite{Taskin2012,YSKim2011,YZhang2010}. Experimental data showed that a Bi$_2$Se$_3$ film has to be at least 5-QL thick to hold the gapless
Dirac cone with a $\pi$ Berry phase. Furthermore, the supercell geometry without vacuum used in Ref.~\onlinecite{ThyTIFM} induces
charge transfer from both neighboring Ni layers to the ultra-thin TI layer. All of these would not favor the formation of the descendent states at
the Fermi level, for the TI-FM layer considered in Ref.~\onlinecite{ThyTIFM}. A slab geometry with a vacuum layer is more relevant to the spin-transfer 
torque experiment\cite{DanExp} than a supercell geometry without vacuum. In our study, we take into account in-plane magnetization used in the 
spin-transfer torque experiment\cite{DanExp}, whereas the latter studies\cite{ThyTIFM,TImetalDFT,TEJA2017} considered out-of-plane magnetization 
due to different experimental set-ups from spin-transfer torque. In our DFT+SOC calculation,
when the magnetization is out of plane, we find that the similar descendent state appears at $E_F$ with the similar
wave function distribution in real space to that shown in Fig.~\ref{fig:DFTbands}(c), and that the descendent state has the 
spin-momentum locking as shown in Figs.~\ref{fig:DFTTIspin}(b) and~\ref{fig:DFTspin}(b).

\section{Conclusion}

We build a simple lattice model in the spirit of Fano-Anderson model and perform first-principles-based simulations in order to address
the fate of the primitive Dirac interface state in TI-FM bilayers under large charge transfer and severe hybridization with many metallic states
overlapping in momentum and energy. Both the tight-binding model and simulations showed that while destroying the Dirac interface state, a large enough hybridization could also create a new descendent state near the Fermi level which inherits both the spatial localization on the interface and the Rashba-type spin-momentum locking. Our findings suggest this hybridization-induced descendent state to be a possible candidate for the source that contributes to the experimentally observed large spin-transfer torque in TI-FM bilayers. While the spin-transfer torque in TI-based structures has attracted growing theoretical attention\cite{MarkTorque,ManchonTorqueTI,ThyTorqueTImetal}, our model provides a starting point for theoretical studies on TI-FM heterostructures to take into account the lowest order effects from the FM layer.
While the hybridization strength is material-dependent, our simple model provides a generic way to describe the hybridization effect for the experimentally relevant cases in which the Dirac interface state overlaps with many FM states.

\noindent
{\bf Acknowledgements} Y-TH was supported by the Cornell Center for Materials Research with funding from the NSF MRSEC program (DMR-1719875) and in part by Laboratory for Physical Sciences and Microsoft. E-AK was supported by U.S. Department of Energy, Office of Basic Energy Sciences, Division of Materials Science and Engineering under Award DE-SC0010313. E.-A.K. acknowledges Simons Fellow in Theoretical Physics Award \#392182 and thanks the hospitality of KITP supported by Grant No. NSF PHY11- 25915.
K.P. was supported by the U.S. National Science Foundation Grant No. DMR-1206354.
The computational support was provided by SDSC under DMR060009N and VT ARC
computer clusters.

%

\end{document}